# Coupling - decoupling of conducting topological surface states in thick Bi$_2$Se$_3$ single crystals

Amit Jash[1], Sayantan Ghosh[1], A. Bharathi[2], S. S. Banerjee[1*]

[1]Department of Physics, Indian Institute of Technology, Kanpur 208016, Uttar Pradesh, India; [2]UGC-DAE Consortium for Scientific Research, Kalpakkam-603104, India;

Sensitive ac-susceptibility measurements of a topological insulator, Bi$_2$Se$_3$ single crystal, using mutual two coil inductance technique (Ref. 32) shows coupling and decoupling of high conducting surface states. The coupling of the surface states exists upto thickness of 70 microns, which is much larger than the direct coupling limit of thickness ~ 5 to 10 nanometers found in thin films. The high conducting topological surface states are coupled through the crystal via high electrically conducting channels, generated by Selenium vacancies. These conducting channels through the bulk disintegrate beyond 70 µm thickness and at high temperatures, thereby leading to decoupling of the topological surface states. We show the decoupled surface states persist upto room temperature in the topological insulator. Analysis of Nyquist plot of ac-susceptibility response of the TI using a resistor (R) – Inductor (L) model shows an inductive nature of the coupling between surface states found in these Bi$_2$Se$_3$ crystals.

## I. Introduction

Three dimensional (3D) Topological insulators (TI) are quantum materials with an insulating bulk enclosed by topologically protected metallic surface states [1,2,3,4,5,6,7,8,9,10,11,12,13] with chiral surface currents [14,15]. Observation of Shubnikov-de Haas (SdH) oscillations in magneto transport measurements [7,16,17] at low temperature (*T*) and observation of the Dirac point located within the bulk gap of the TI by ARPES [15,18] confirm the 2D topological nature of the surface state. Recent studies in thin film [19,20,21,22,23,24] show that for thickness below a critical value ~ 6 nm (direct coupling limit), the top and bottom topological surface state (TSS) wave functions overlap and hybridize. This leads to a disappearance of the Dirac cone in TI [19] by opening a gap at the surface state [25]. The direct coupling of TSS was shown to be capacitive in nature in thin films [23,24]. The above implies that for TI materials with thickness greater than 6 nm, the conducting TSS states should decouple. A question naturally arises that in 3D TI's does coupling-decoupling exist? Another related question is, does decoupled high conducting TSS survive upto room temperature in TI's? While these issues have been extensively investigated for films, they have not received much attention in 3D TI's. Furthermore, it may also be mentioned that evidence of high conducting TSS at room $T$ is important for developing applications using TI's [26,27].

---
[*] Email: satyajit@iitk.ac.in.

Most 3D TI's always contain some residual bulk carriers that couple to surface states through disorder and phonons [28,29,30,31]. In 3D TI material $Bi_2Se_3$, the presence of atomic vacancies like selenium (Se) vacancies, dopes the bulk which leads to opening of parallel conduction channels through the bulk. In this situation, transport measurements cannot distinguish between bulk and surface contributions to conductivity, especially at higher $T$. Recently, a non-contact two coil mutual inductance measurement technique [32] separately identified the surface and bulk contributions to conductivity in TI. Here, using this technique we explore the ac susceptibility response of the TSS in $Bi_2Se_3$ single crystals as a function of temperature and sample thickness ($d$). Frequency dependence of the mutual inductance which is proportional to the ac-susceptibility, shows a characteristic frequency $f_0$ where conduction transforms from bulk to surface. The $f_0(d)$ behavior is modelled with high conducting channels coupling the TSS. We show coupling is established even at thickness larger than the direct coupling limit. Study of in-phase and out of phase component of the mutual inductance shows the presence of TSS in TI at room $T$. Beyond a critical thickness of ~ 69 μm and near room $T$ these coupling channels disintegrate, leading to a decoupling of the TSS and producing high conducting TSS at room $T$. Nyquist plot of the data and its modeling based on resistance–inductance network identified the inductive coupling through the TI.

Single crystals of $Bi_2Se_3$ are prepared by slow cooling stoichiometric melts of high purity bismuth (Bi) and selenium (Se) powders [33, 34]. We study five different $Bi_2Se_3$ single crystals from the same batch with thickness 20 μm (S20), 49 μm, 69 μm (S69), 75 μm (S75), 82 μm (S82). Average mobility of the samples is about 3000 cm$^2$/V-s. We use two - coil mutual inductance setup [32] to measure the ac-susceptibility response of the TI.

## II. Experimental Setup:

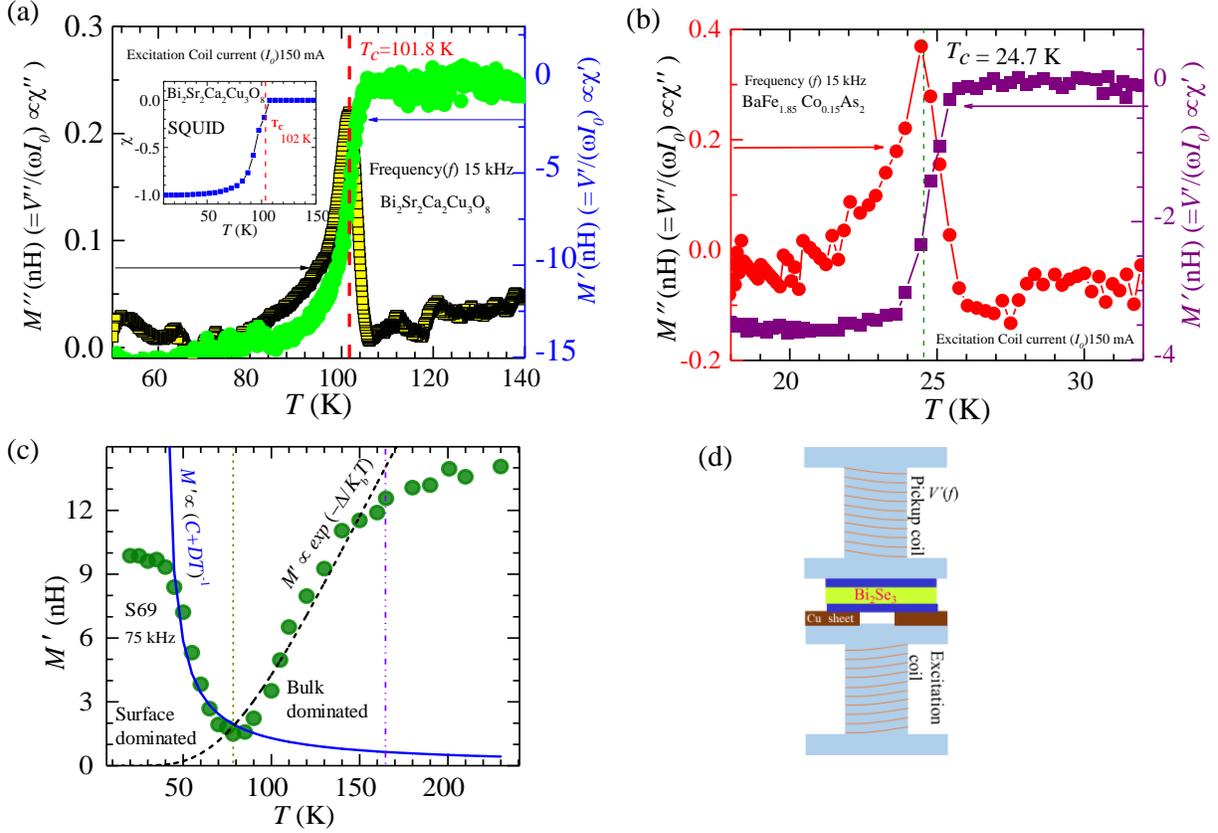

Fig. 1. Shows the real ($M'$) and imaginary ($M''$) component of the mutual inductance as function of temperature for (a) high temperature Cuprate superconductor ($Bi_2Sr_2Ca_2Cu_3O_8$), (b) Pnictide superconductor ($BaFe_{1.85}Co_{0.15}As_2$). Inset of figure (a) shows ac susceptibility as measured on a Cryogenics SQUID magnetometer. (c) shows the $M'(T)$ response of $Bi_2Se_3$, for details see text. (d) shows the schematic of two-coil setup.

The two-coil mutual inductance measurement setup is used as a sensitive non-contact technique to distinguish between the surface and bulk conductivity of a TI [32]. The schematic of the two-coil setup is shown in the Fig. 1(d). In this technique the TI sample to be investigated is placed between a primary excitation coil and secondary pickup coil. A small ac signal is applied across the excitation coil. The excitation current generates a time-varying magnetic field, which induces currents inside the conducting sample, which in-turn produces a time-varying magnetic field which induces a voltage in the pickup coil. Effectively one considers that the presence of a sample between the two coils modifies the mutual inductance and changes in the sample properties changes the signal induced in the pickup coil. To enhance the coupling of the signal between excitation and pickup coil via the sample, and to reduce the stray field coupling between the coils, a 100 µm thick sheet of oxygen free high purity copper (Cu) with a hole, is placed between the sample and the excitation coil. The hole diameter is chosen to be just less than the sample width. This Copper sheet with hole serves multiple purposes. The thickness of the Cu sheet is chosen to be larger than the skin depth of the ac field produced by the excitation coil. Thus, outside the borders of the sample the Cu sheet minimizes the stray field coupling between the excitation and pickup coil, as these fields are shielded by the Cu sheet. Within the hole in the Cu sheet, the magnetic flux is focused onto the sample which is placed directly above the hole [32]. The above ensures that maximum flux linkage happens between the excitation and pickup coil through the sample rather than through the stray magnetic fields outside the sample (which are shielded by the 100 µm Cu sheet). This in turn makes the pickup signal sensitive to detecting small change in the conducting properties of the sample. It has been demonstrated that using this technique, there is almost two times increase in the pickup signal from the sample [32]. Note that for all our measurements we subtract the background voltage at all temperature and frequencies by measuring the pickup signal without a sample and with only the Cu sheet with hole placed between the coils as the background signal. The background signal is measured at each temperature and ac excitation frequency using the Copper sheet with hole placed between the coils (with no sample present between the coils).

The mutual inductance of the sample which couples the excitation and pickup coils signals is measured in this technique via the pickup voltage $V$, mutual inductance $M = \frac{V}{\omega I_0}$, where $\omega$ is the angular frequency of the excitation signal in the primary coil and $I_0$ is the amplitude of the ac current in the primary coil. The sample's mutual inductance and its ac susceptibility response ($\chi_{ac} = \chi' - i\chi''$) are related: The voltage induced in the secondary coil (i.e., the pickup voltage) is $V = M\frac{di_{ac}}{dt}$, then $V = -Nk\chi_{ac}\frac{dh_{ac}}{dt}$, where $h_{ac}$ is the field by the primary coil, $N$ is the nos. of turns in the secondary coil and $k$ is the geometric filling factor. Thus, the mutual inductance ($M$) and ac susceptibility are directly related. We compare the $T$ dependent behaviour of the real ($M'$) and imaginary ($M''$) component of the mutual inductance as deduced from the in phase and out of phase components of the pickup voltage ($V'$ and $V''$) in Fig. 1(a) for the high temperature superconductor BSCCO-2223 ($Bi_2Sr_2Ca_2Cu_3O_8$) with a $T_c$ of 101.8 K, Fig. 1(b) Pnictide superconductor ($BaFe_{1.85}Co_{0.15}As_2$) with $T_c$ of 24.7 K. The data in these figures are obtained after background subtraction. It is worthwhile noting that a comparison of Figs.1(a) to 1(c) shows the sample response are different for different sample. In Figs. 1(a) and 1(b) the real and imaginary parts of mutual inductance ($M'$) and ($M''$) are shown as a function of temperature. The green data points in Fig. 1(a) and the purple data points in Fig. 1(b) show the real part of mutual inductance ($M'$) which is proportional to $\chi'$. These figures show the expected behaviour of $\chi'(T)$ of a superconductor associated with the development of a large diamagnetic shielding response as the temperature falls below $T_c$. The yellow data in Fig. 1(a) and red data in Fig. 1(b) shows the behaviour of the imaginary component of $M$ which is proportional to $\chi''$, showing the characteristic dissipation peak develop near $T_c$ of a superconductor. All these features show that our system sensitively measures the ac susceptibility response of a sample. For comparison of our data with that from another standard measuring instrument, in the inset we show the ac susceptibility response of the same superconductor as measured on a Cryogenics SQUID magnetometer. The similarity of the two in phase ac-susceptibility measured establishes we are measuring the $\chi'$ response of a superconductor.

Using temperature dependent mutual inductance measurement, we identify the distinct temperature dependence of surface conductivity and bulk conductivity in Bi$_2$Se$_3$ [32]. The $M'(T)$ data in Fig. 1(c) shows that the mutual inductance saturates to constant value below 40 K (such a feature is well known for TI material, Bi$_2$Se$_3$ where the surface conductivity saturates at low $T$). Above 40 K to 70 K the $M'(T)$ data fits (solid blue line) to a form $1/(C+DT)$ where $C$ is related to static disorder scattering and $D$ is electron-phonon coupling strength. This form is associated with the conduction of surface state in the TI [32]. Above 70 K to 170 K the data fits (red dashed line) with $exp(-\Delta/K_bT)$, $\Delta \sim 25.2 \pm 1.25$ meV, where $\Delta$ is an activation energy scale. This thermally activated behavior corresponds to $M'(T)$ response associated with bulk conductivity. The thermally activated nature is attributed to thermal activation of charges from disordered sites in the bulk of Bi$_2$Se$_3$ [32].

Although, we perform a background subtraction, of the Cu sheet with a hole, the signal from the Cu cannot be completely eliminated. However, as we show below this contribution is very weak compared to the sample signal. In BSCCO (Fig. 1(a)) for example, well below $T_c$ one expects the combined response of the Cu sheet with hole along with the superconducting contributions to the net signal. However, above $T_c$ the measured signal would be primarily only due to the Cu sheet with hole, as susceptibility response of the superconductor is vanishingly small in the normal state ($T > T_c$). Hence, the normal state signal (above $T_c$) is compared with the total diamagnetic signal below $T_c$ to figure out the extent of error contributed to the signal from the Cu sheet with hole. The percentage error $\varepsilon$ in the signal due to the Cu sheet with hole is estimated as $\varepsilon \sim \frac{<M'(T>T_c)>}{|M'(15K)-<M'(T>T_c)>|} \times 100$, where <..> represents the mean value. From Fig. 1(a) we estimate $\varepsilon$ =1.7 % for BSCCO and for BaFe$_{1.85}$Co$_{0.15}$As$_2$ (Fig. 1(b)) $\varepsilon$ = 2.07 %. Hence, after background subtraction procedure the parasitic influence of the focusing copper strip is minimal and far less than the sample signal. Therefore, from this error estimation we believe that the response we have measured is essentially that of the sample with negligible contributions from the focusing Cu sheet with hole. Hence in two-coil technique with the copper sheet with hole present for focusing the signal onto the sample, 98 to 99 % of the signal measured is from the sample being explored and the remaining 1 to 2 % of the signal is from the Copper sheet. This 1 to 2 % contribution from Copper isn't strong enough to overwhelm and dominate over the distinct nature of the signals measured from superconductors (Figs .1(a) and 1(b)) and topological insulator (Fig. 1(c)). Had the Cu sheet contributed significantly to the signal, this difference could not have been observed.

### III. Result and discussion:

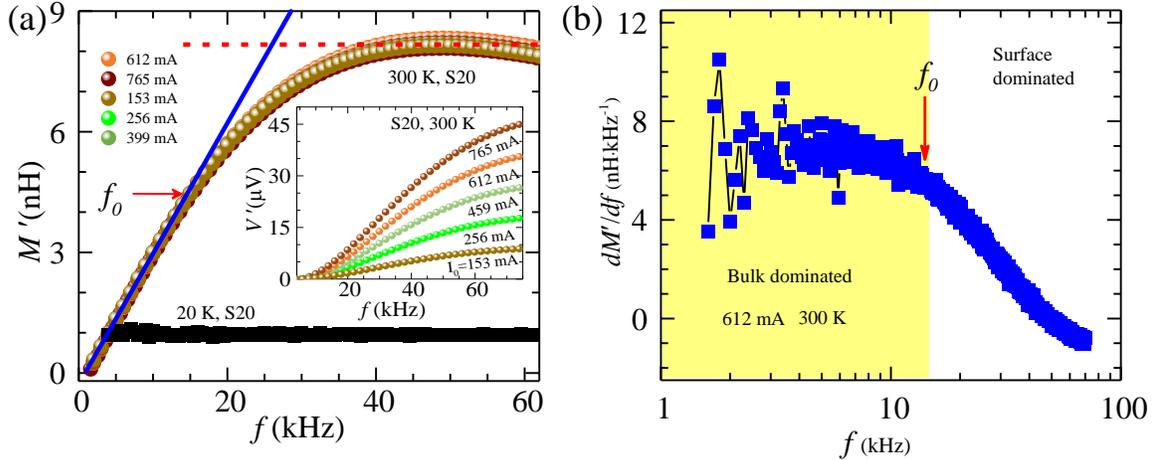

Fig. 2. (a) scaled real part of mutual inductance, $M'(f)(= \frac{V'}{I_0 2\pi f})$ at 300 K and $M'(f)$ at 20 K (black square) while inset shows $V'(f)$ for different ac excitation current amplitudes. (b) shows the $dM'/df$ versus $f$ where $f_0$ is calculated from the constant regime (which is shaded as yellow).

It is well known that for any conducting sample the skin depth is proportional to $1/\sqrt{f}$ where $f$ is the frequency of an impinging ac field [32]. Hence in our two-coil setup with a TI placed between the coils, at low frequencies ($f$) the signal would penetrate deeper into the sample resulting in the shielding currents being induced throughout the bulk of the sample. Due to this in the two-coil setup at low $f$ one expects to measure the average bulk response of the material (proportional to the bulk ac susceptibility response) of the TI [32]. At higher $f$ as the penetration depth of the signal decreases, one begins to probe features closer to the surface of the TI. Hence at high $f$, currents are induced primarily near the TI sample surface. Therefore, at high $f$ the behavior of the surface conductivity of sample is effectively probed [32]. Hence, using this two-coil mutual inductance technique for TI's, by using excitation signals of different $f$ the bulk and surface conductivity response of a TI can be probed separately. This is unlike conventional transport measurements where due to parallel conducting channels between the bulk and surface of the TI, the conductivity response of surface and bulk cannot be clearly distinguished through transport measurements. It has been shown [32] that for surface dominated conduction in TI's, $V' \propto f$ or $M'(\equiv$ real part of ac-susceptibility, $\chi'$) is constant and when the conduction is bulk dominated then $V' \propto f^2$ or $M' \propto f$. Inset of Fig. 2(a) shows $V'(f)$ measured at 300 K for different $I_0$, where all curves collapse onto a single $M'(f)$ curve in Fig. 2(a). Figure 2(a) also shows a linear $M' \propto f$ regime up to 14 kHz (bulk dominated), and $M' =$ constant regime beyond 40 kHz (surface dominated) at high $T$. Similar to the S20 sample, other samples also show same behavior (see supplementary information, section SI1). The above suggests that at high $T$, bulk contribution mixes with surface dominated conductivity. In Fig. 2(a), the linear frequency dependent $M'(f)$ transforms to a frequency independent regime near 14 kHz. This characteristic frequency $f_0$ has been identified in Fig. 2(b) where $dM'/df$ is plotted with $f$. Recall that at 15 K, SdH oscillations [32] showed the dominance of 2D like surface conductivity in the TI (see SI2). In Fig. 2(a) the $M' =$ constant at all $f$ at 20 K shows the dominance of TSS conduction over the bulk at low $T$.

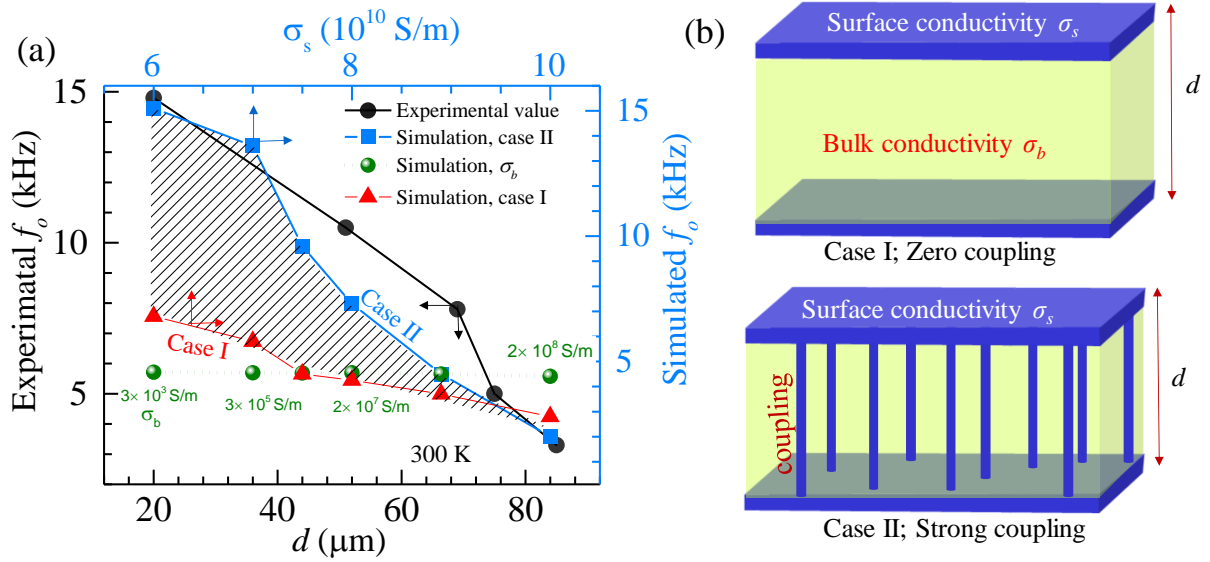

Fig. 3. (a) measured $f_0(d)$ behavior (black circle symbol) behavior (left axes) at 300 K and simulated $f_0(d)$ (right axis) (blue square symbol, case II). The red line (triangle symbol) shows simulated $f_0(d)$ with zero coupling (case I). The green circle data points show simulated $f_0(d)$ when the bulk conductivity is varied keeping surface conductivity fixed. (b) Case I (zero coupling scenario): shows the schematic of TI where blue region is surface state with conductivity $\sigma_s$ and yellow region represents the bulk with conductivity $\sigma_b$. Case II (coupled scenario): where the conducting sheets are connected via the blue high conducting channels with conductivity $\sigma_s$ in the bulk (blue channel).

Using analysis like in Fig. 2(b), we determine the behavior of $f_0(d)$, which is shown in Fig. 3(a). The experimental data in Fig. 3(a) shows $f_0$ decreases with $d$ at 300 K. To understand this $f_0(d)$ behavior, we simulate the electromagnetic response of TI when oscillating magnetic fields impinge upon the material. The TI is modelled as a relatively low conducting material with bulk conductivity $\sigma_b$ sandwiched between two highly conducting thin sheets (viz., TSS) of thickness 10 nm with conductivity $\sigma_s$ (see Case I in Fig. 3(b)). The simulation of $M'(f)$ of such a sample placed between excitation and pickup coils used in our experiments is carried out using COMSOL Multiphysics software(see section SI3 and section $S2$ of ref.[32]). The simulations (see SI4 for details) also show the linear and constant $M'(f)$ regimes from which the simulated $f_0$ is determined. In the simulations by keeping a fixed $\sigma_s = 6 \times 10^{10}$ S/m, $\sigma_b = 3 \times 10^3$ S/m and varying $d$ we do not observe any significant change in simulated $f_0$ value from 7 kHz using case I model while experimental $f_0$ value (black circle) deviate from this. Again, by keeping $d$ and $\sigma_s$ fixed and varying $\sigma_b$ there is no appreciable change in $f_0$ value (see green circles). We find that the simulated $f_0(d)$ approaches close to the experimental behavior by using the Case II model of the TI sample. In case II model the high conducting surface sheets are coupled through the bulk via the high conducting channels also with electrical conductivity $\sigma_s$ [32]. $Bi_2Se_3$ has Se vacancies which lead to electron doping in its bulk [33,35,36]. We model through Case II the excess conducting charges emanating in the bulk due to Se vacancies, as depicted by the high conducting cylinders (Fig. 3(b)). Figure 3(a) shows a good match between simulated and experimental $f_0(d)$ by using $d = 20$ µm, $\sigma_b \sim 5 \times 10^3$ S/m and varying $\sigma_s$ from $6 \times 10^{10}$ S/m to $10^{11}$ S/m. Thus, by varying $d$ there are some fundamental changes in the TI which correspond to enhancing the surface conductivity $\sigma_s$. At higher $d$ (beyond 69 µm), both the experimental $f_0(d)$ and simulated $f_0$ calculated using case I and case II model merge together. This suggests that at high $d$ the role of coupling between the metallic sheets diminishes. Here the surface conductivity of the sheets is higher than that at lower $d$ (< 69 µm) where the conducting TSS sheets are strongly coupled. We believe at lower thickness the conducting sheets are coupled through the high conducting channels permeating the bulk. At higher thicknesses these conducting channels get disrupted and become discontinuous, thereby decoupling the TSS. From Fig. 2, the frequency independent $M'(f)$ at high $f$ at 300 K confirms the presence of a high conducting TSS persisting at room $T$.

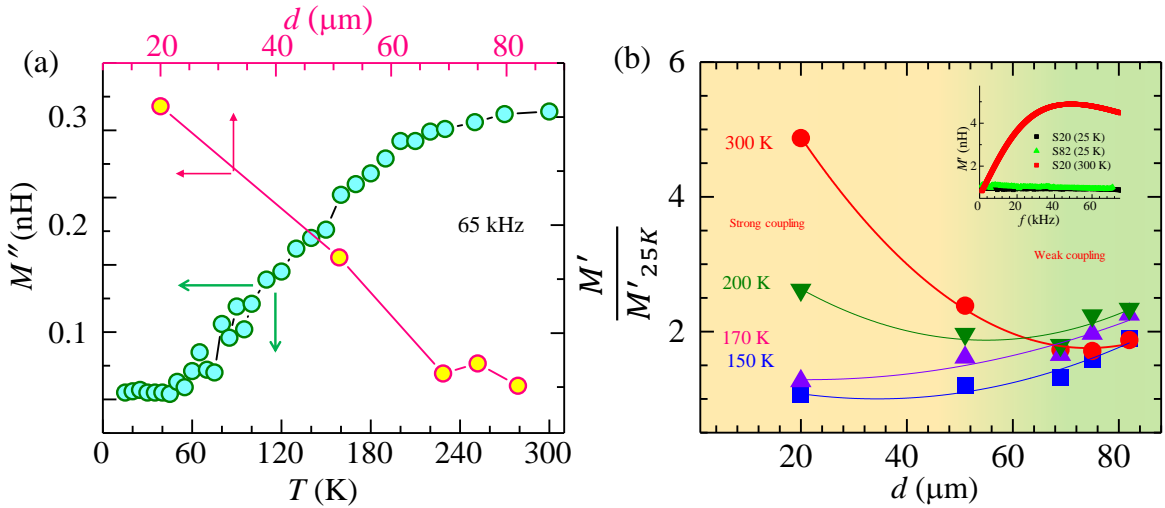

Fig. 4. (a) $M''$ vs $T$ for S20 sample (left axis) and the right axis shows how $M''$ vs $d$ (top axis) at 300 K. (b) shows $\frac{M'}{M'_{25K}}$ as function of sample thickness for 150 K, 170 K, 200 K and 300 K. Inset shows the $M'(f)$ at 25 K for S82, S20 and at 300 K for S20.

Figure 4(a) shows the behavior of the measured imaginary component of $M$, viz., $M''$ ($\propto \chi''$, imaginary component of ac-susceptibility) which is a measure of dissipation in the material. The $M''(T)$ expectedly shows an increase in dissipation with increasing $T$ and saturates above 250 K. We believe that the TSS do not contribute to dissipation as the chiral surface currents do not scatter. Figure 1(c) shows the thermally activated conductivity of charges created by Se vacancies begins from 70 K. To delocalize these charges an activation barrier in the range of 10 to 25 meV [32] is overcome at ~ 70 K. Due to this, at $T$ below 70 K there is minimal contribution to bulk conductivity from these doped electrons and the conductivity is dominated by TSS. Hence the dissipation at low $T$ is relatively low and at a constant level. Thus, the dissipation arises from the bulk conducting channels which couple the surface states (Case II of Fig. 3(b)). Figure 4(a) also shows (upper axis) that $M''$ (at $T$ = 300 K) decreases with increasing $d$, and the dissipation is lowest and becomes constant for, $d$ > 69 µm. The observed decrease in dissipation with increasing thickness is due to disintegration of continuous conducting channels present through the bulk which leads to a decoupling of the TSS. We believe in these bulk single crystals of $Bi_2Se_3$ decoupling of the TSS occurs close to 70 µm.

Below 25 K due to domination of surface conductivity [9,10,13,32], the $M'$ is independent of $f$ (see Fig. 2(a), and inset Fig. 4(b)). The inset of Fig. 4(b) also shows that at higher $T$ the $M'$ develops a frequency dependence, due to the contribution from bulk conductivity. In Fig. 4(b) we plot the behavior of $M'(d)$ at different $T$. Note that all $M'$ values are normalized by the value of $M'$ at 25 K, viz., the $T$ where conductivity is dominated by the TSS. At 150 K and 170 K the $M'/M'_{25K}$ begins from a value slightly above 1 for S20 and then monotonically increases with $d$. Note that for 150 K and 170 K the $M'/M'_{25K}$ ~ 1 does not mean at these temperatures all conductivity is surface dominated. Rather, it represents that at higher $T$ due to the increasing bulk contribution to conductivity the $M'$ value is similar to the value at 25 K (Fig. 1(c)). This results in $M'/M'_{25K} \geq 1$ with increasing $T$. Here we draw attention to the behavior of spread in the $M'(T)/M'_{25K}$ values as a function of $d$. Note that above 69 µm the $M'/M'_{25K}$ behavior becomes almost $T$ and thickness independent. At a $d$ = 20 µm the large spread in the $M'/M'_{25K}$ value for different $T$ in Fig. 4(b) corresponds to the enhancing contribution to bulk conductivity, as 150 K is well above the activation barrier (70 K) required to delocalize the doped electrons from Se vacancy centers in the $Bi_2Se_3$ lattice. These activated electrons migrate from the bulk to the surface due to band bending effects [37,38], thereby creating the high conducting coupling channels to the surface states in the relatively thinner samples (case II in Fig. 3(b)). Although in all the samples the thermally activated doped electrons are generated, in thicker samples with surfaces further apart, the band bending has less effect on the doped electrons in the bulk. Hence the conducting channels disintegrate in thicker samples leading to the $M'(T)/M'_{25K}$ values becoming $T$ and $d$ independent. In the absence of conducting channels, the net effective contribution to the electrical conductivity from the bulk saturates and hence the $M'/M'_{25K}$ value becomes almost $T$ independent for samples thicker than 69 μm. With the conducting channels in the bulk disintegrating for larger $d$, the conducting surface states in the 3D TI consequently get decoupled. Recall that Fig. 3(a) showed the merge of experimental and simulated $f_0(d)$ for $d \geq 69$ μm due to decoupling transition of TSS. Figure 3(a) shows the experimental $f_0(d)$ value simulated successfully for lower $d$ with lower $\sigma_s$ while for the thick sample $\sigma_s$ is at least an order of magnitude larger. We propose that the bulk conducting channels may be introducing a coulombic drag which reduces the surface conductivity in thinner samples. In thicker samples $\sigma_s$ increases with weakening of the conducting channels.

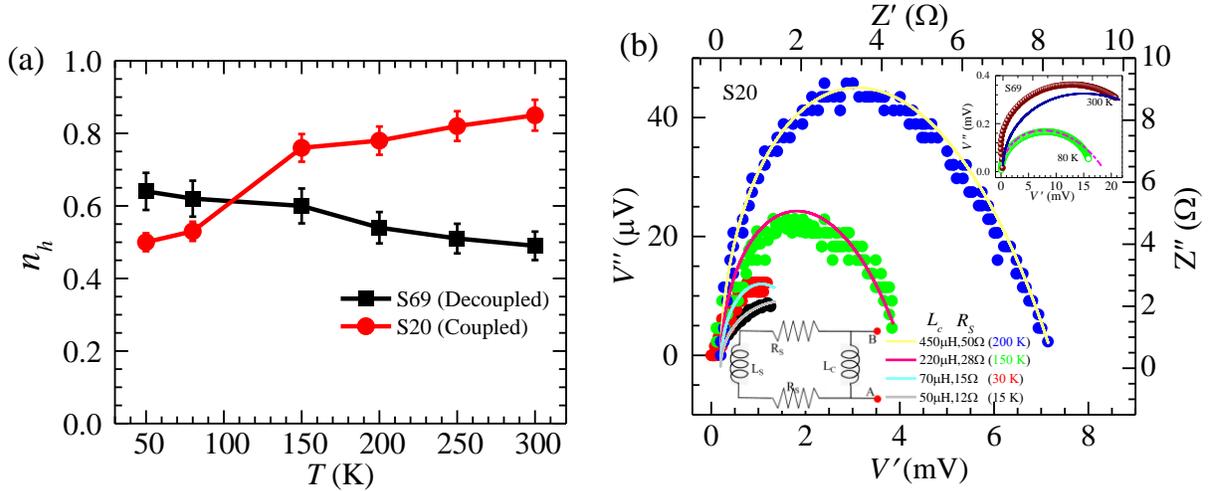

Fig. 5. (a) the high conducting fraction $n_h$ vs $T$ for S20 and S69 samples. (b) Nyquist plots for S20 at various temperatures. Top inset, Nyquist plots for S69 at 80 K and 300 K. Bottom panel is the equivalent inductance ($L$) and resistance ($R$) circuit model.

In Fig. 5(a) we estimate the temperature dependence of the high conducting fraction ($n_h$) of the TSS and the high conducting channels in the bulk. We estimate $n_h(T)$ for two samples, viz., the S20 (coupled regime) and S69 (uncoupled regime) samples. From Fig. 2(a) we know at low $f$, viz., $f < f_0$, $M'(f) = a.f$ while for higher $f$ ($> f_0$), $M'(f) \sim b$ where $a$ and $b$ are constants, which are determined from the slope of the $M'(f)$ data at low $f$ ($< f_0$) and from the saturated value of $M'(f)$ at high $f$ ($> f_0$) respectively. Finally the $M'(f,T)$ is fitted using $M'(f) = n_L.a.f + n_h.b$, where $n_L$ is the low conducting fraction of the TI sample (by volume) with the constraint that $n_h + n_L = 1$. Figure 5(a) shows that the variation in $n_h(T)$ is almost temperature independent for the thicker uncoupled S69 sample while comparatively there is a significant change in $n_h(T)$ for the coupled S20 sample. As the high conducting channels disintegrate in the thicker S69 sample, hence the TSS are decoupled and therefore the $n_h(T)$ remains roughly constant upto high $T$ of 300 K. This feature also confirms our previous finding of the TSS persisting up to room $T$, where recall from Fig. 5(a) we observe high conductivity at room $T$ with Figs. 2 and 4 showing presence of surface conduction at room $T$. Whereas in S20 sample due to thermally activated delocalization of the Se vacancy doped charges, newer conducting channels are created which lead to an enhancement in $n_h(T)$. In Fig. 5(b) and in the inset we show the Nyquist plots, viz., the out of phase component of the pickup signal viz., $V''(\propto \chi'')$ plotted against the in phase component, viz., $V'(\propto \chi')$, at different $T$ for S20 (coupled) and S69 (decoupled) samples, respectively. To understand these plots, we introduce a simple resistor ($R$) – inductor ($L$) network model. The TSS states in the TI are modelled with two resistors $R_S$ and $L_S$ (see schematic in Fig. 5(b)). The resistors $R_S$ represents the resistances of the surface states on the top and bottom of the TI. The inductor $L_C$ represents the coupling between the surface states in the TI through the bulk, viz., via the conducting channel present in the bulk. Note, $R_S$ includes the resistance of the coupling channels. For this model we calculate the ac impedance between points A and B as $z' = \dfrac{2R_s \omega^2 L_s^2}{4R_s^2 + \omega^2 (L_s + L_c)^2}$ and

$z'' = \dfrac{4L_s \omega R_s^2 + L_c L_s \omega^3 (L_s + L_c)}{4R_s^2 + \omega^2 (L_s + L_c)^2}$. Using these expressions and a constant $L_S \sim 2.5$ µH the behavior of $Z''$ versus $Z'$ for different $R_s$ and $L_c$ are plotted with solid lines in Fig. 5(b) and inset. The curvature of $Z''(Z')$ curve follows closely the $V''$ vs $V'$ curve. As Fig. 5(b) shows that the theoretical and experimental curves match (qualitatively) only if the coupling inductance $L_C$ is increased as $T$ is increased for the S20 sample. In fact, from 50 K to 200 K the coupling $L_C$ increases by almost an order of magnitude. This is consistent with what we found earlier, viz., the S20 sample gets more strongly coupled due to opening of newer conducting channels with increasing $T$. However, in S69 (inset of Fig. 5(b)) we see that only at low $T$, $Z''(Z')$ curve follows the $V''$ vs $V'$ curve. At higher $T = 100$ K and above the coupling, inductance is unable to fit the curvature of $V''$ vs $V'$ curve. This suggests that in thicker samples like S69 although conducting channels may open, they are not continuous and therefore the top and bottom conducting surfaces decouple. We also must clarify that at the very low $T$ where the probability of thermally activated charges in the bulk is very low, here also for our thinnest S20 crystal, it is thicker than the direct coupling limit (~ 6 nm). Therefore, the TSS in these crystals are in a decoupled state at low $T$. The coupling of TSS develops in these crystals above 70 K when there is enough activation of bulk charge carriers which couple TSS via the conducting channels. Whereas in the direct coupled regime in thin films, at few nm there is a quantum mechanical hybridization of the surface states which leads to a destruction of the TI state [25], but here for the 3D no effects happen. It was reported in past that nature of coupling between the TSS is capacitive [23,24] in 2D thin film. Here we observe inductive coupling between the TSS in a 3D TI. While we do not know the mechanism of this coupling, we believe it leads to the conducting surfaces behaving as a coupled entity with a relatively lower surface conductivity when coupled at lower $d$ than when they are decoupled at higher $d$, while preserving the TI nature up to room $T$.

## IV. Conclusion:

In conclusion we have reported a novel coupling – decoupling transition in thick single crystals of $Bi_2Se_3$. The phenomenon is disorder driven and arises as a result of Se vacancy induced electron doping in the bulk of $Bi_2Se_3$ single crystals at temperatures close to 70 K. We find that at high $T$ the disorder effect leads to coupling of the TSS to persist up to large thickness which are much greater than the direct coupling limit. Our studies show the persistence of a high conducting TSS up to room $T$ in $Bi_2Se_3$ single crystal. These studies we believe are useful for exploiting the topological features of $Bi_2Se_3$ for room temperature applications. We believe this disorder driven coupling-decoupling phenomenon needs further theoretical and experimental investigations. Our evidence of survival of high conducting surface states at room $T$ is also useful from the point of view of applications and needs further exploration.

**Acknowledgment**

SSB acknowledges funding support from DST (AMTTSDP and Imprint-II programs), IIT Kanpur and help of T. R. Devidas from IGCAR Kalpakam. SG thanks CSIR for funding support.

# Supplementary information

**Coupling - decoupling of conducting topological surface states in thick $Bi_2Se_3$ single crystals**

Amit Jash[1], Sayantan Ghosh[1], A. Bharathi[2], S. S. Banerjee[1*]

[1]Department of Physics, Indian Institute of Technology, Kanpur 208016, Uttar Pradesh, India; [2]UGC-DAE Consortium for Scientific Research, Kalpakkam-603104, India;

## SI1: S69 sample response:

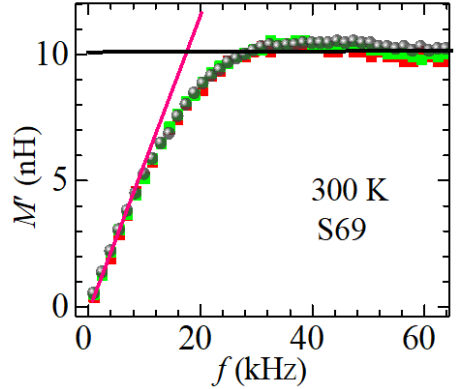

Figure above shows the scaled behaviour of the pickup voltage of S69. Scaling is done by using this form, $V' = I_0 M' \omega$ where $I_0$ is the amplitude of the excitation signal. The $M'(f)$ profile for S69 is almost similar with S20 sample (Figure 2(a) in main manuscript), only the $f_0$ value is much smaller than S20 sample. Figure shows two regions distinctly, linear region is shown by red solid line and saturation region is shown by black solid line. After 9.5 kHz, $M'$ deviates from the linear region. Other samples also show similar behaviour.

## SI2: SdH oscillation:

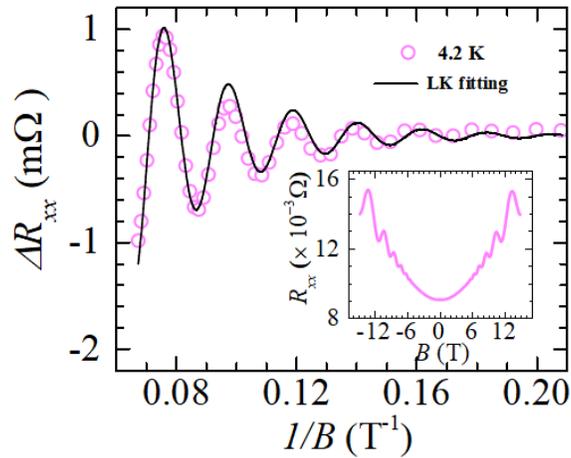

Figure 1 shows distinct SdH oscillation in longitudinal magneto-resistance ($R_{xx}$ vs magnetic field ($B$)) measurements at 4.2 K using standard Van der Pauw geometry (reproduced partially from Fig. 1(a) in Ref. [38]). Inset shows the variation of $R_{xx}$ as a function of magnetic field $B$ measured at 4.2 K. The main panel shows $\Delta R_{xx}$ vs. $1/B$ plot showing SdH oscillations in $Bi_2Se_3$ (sample thickness 70 μm) at different temperatures. $\Delta R_{xx}$ is calculated by subtracting from the experiment $R_{xx}(B)$ values a polynomial fit to the data ($R_{poly}(B)$), i.e, $\Delta R_{xx} = R_{xx}(B) - R_{poly}(B)$. The polynomial form of $R$ is $R_{poly}(B) = R_0 + R_1 B + R_2 B^2$, where $R_0 = 9.26 \times 10^{-3}$ Ω, $R_1 = -1.71 \times 10^{-6}$ Ω·T$^{-1}$, $R_2 = 2.91 \times 10^{-5}$ Ω·T$^{-2}$. Lifshitz-Kosevich (LK) equation is used to fit the $\Delta R_{xx}$ data at 4.2 K which is shown by black solid line. For fitting to the magneto-resistance data in Fig. 1(a), we use $\Delta R_{xx} = a\sqrt{0.011B} \left( \frac{\frac{11.12}{B}}{\sinh\left(\frac{11.12}{B}\right)} \right) e^{-\frac{19.38}{B}} (0.95) \cos\left[ 2\pi \left( \frac{F}{B} + \beta \right) \right]$. In this equation $a = 0.0135$ Ω and $F$ and $\beta$ are the fitting parameters. A fit to the data (see black solid line in Fig. 1(a)), gives $F = 46.95 \pm 0.25$ T and $\beta = 0.44$. The Berry phase ($\phi$) is calculated from the LK fitting and the value being close to π at low $T$, suggests the SdH oscillation arises from surface Dirac electron and charge density calculated from the fitting parameter ($F$) does not match the bulk charge density.

## SI3: COMSOL simulation.

We have also verified our experimental data with simulation using a simple model. The simulation part has been performed using Comsol multiphysics software (AC-DC module). The simulation is done by the solving the standard Maxwell EM equations:

$$\left( j\omega\sigma - \omega^2 \varepsilon_0 \varepsilon \right) \vec{A} + \vec{\nabla} \times \frac{\vec{B}}{\mu_0 \mu_r} - \sigma \vec{v} \times \vec{B} = \vec{J}_e$$

$$\vec{\nabla} \times \vec{A} = \vec{B}$$

$$J_e = \frac{N(V_{coil} + V_{in})}{R_{coil}}$$

where $\omega$ is the angular frequency of the applied AC signal, $A$ is the magnetic vector potential, $\sigma$ is the conductivity of the material, $v$ is the velocity of the charge particle, N is the number of turns, $R_{coil}$ is the resistance of the coil, $J_e$ is the current density and $V_{coil}$ is the applied AC voltage in the coil. The overall schematic of the Comsol simulation is shown in Fig. (a). Figure (b) shows the schematic of ideal topological insulator. The sample is 20 μm (with bulk conductivity $\sigma_b$) thick and the two surface states (with conductivity $\sigma_s$) are 10 nm thick each. Figure (c) shows the schematic of inhomogeneous bulk state. The coupling channels are the cylinders having diameter 100 nm. For the case I, there is no coupling channels between the surface states as $\sigma_s$ is varied keeping $\sigma_b$ constant. For case II, both surface states and coupling channels conductivity is varied from $6 \times 10^{10}$ S/m to $1.2 \times 10^{11}$ S/m keeping $\sigma_b$ constant.

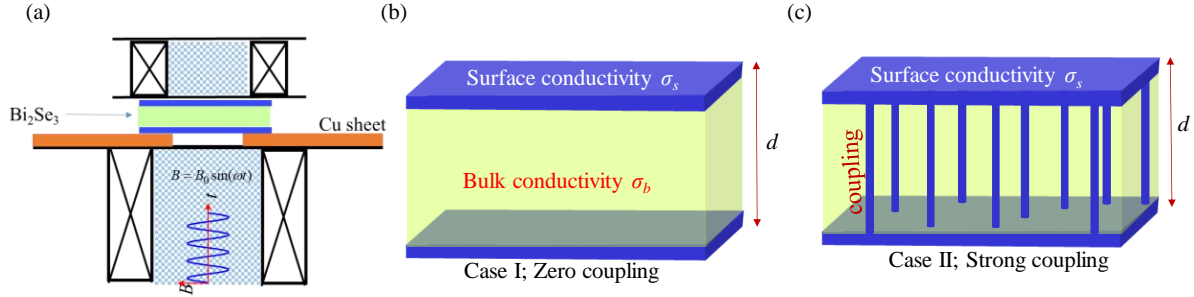

**SI4: Simulation result, $f_0$ for surface and bulk conductivity:**

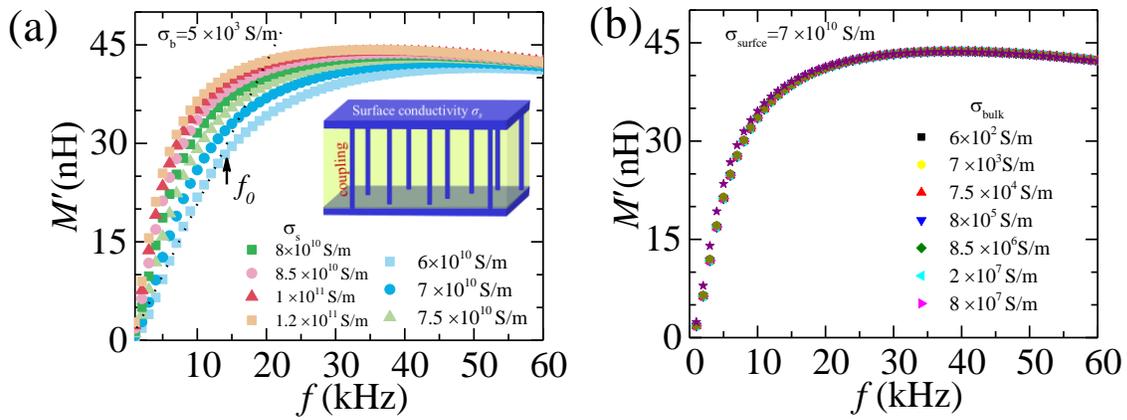

Figure (a) shows the $M'(f)$ behavior for different surface conductivity keeping bulk conductivity constant. Surface conductivity is changed from $6 \times 10^{10}$ S/m to $1.2 \times 10^{11}$ S/m. Using our Inhomogeneous topological model (shown in the Fig. 3(b)), $M'(f)$ is calculated. We can observe that saturation region appears at lower frequency when the surface conductivity is higher which indicates $f_0$ can be tuned with surface conductivity. Figure (b) shows that $M'$ as function of frequency for different bulk conductivity where surface conductivity is kept constant. Bulk conductivity has no direct role in $f_0$. Though the bulk conductivity is changed from $6 \times 10^2$ S/m to $8 \times 10^7$ S/m still the $M'(f)$ profiles are similar.